# Effect of Cavitating-bubble Generators on the Dynamics of Unsteady Cloud Cavitation

Ebrahim Kadivar[1], Khodayar Javadi[2]

[1]University of Duisburg-Essen, Institute of Ship Technology, Ocean Engineering and Transport Systems;
[2]Sharif University of Technology, Department of Aerospace Engineering, Flow Control Research Lab.

**Abstract**
The present paper is devoted to the 3D numerical study of unsteady cloud cavitation around the CAV2003 benchmark hydrofoil using a passive cavitation controller method called cavitation-bubble generator (CGs). We investigate this passive controller on the hydrofoil surface to affect the whole processes of vaporization, bubble generation and bubble implosion. Firstly we simulate the unsteady cavitating flow around the hydrofoil without CGs using a Partially-averaged Navier Stokes (PANS) method. The turbulence approach is coupled with a mass transfer model which implements to an open source finite-volume package. Then we tested different sizes and angles of CGs on the hydrofoil surface. The results of an appropriate design of the CGs shows a significant decrease of the maximum length of the attached cavity and a large decrease of wall-pressure fluctuation intensity on the hydrofoil surface.

**Keywords:** Cloud cavitation, passive flow control, Cavitation Control

**Introduction**

The passive and active flow control methods has long been used in order to control and reduce undesirable behaviors of cavitation. Crimi et al. [1] investigated the effect of introducing a sweep angle to a hydrofoil. They carried out tests on semi-span hydrofoils with sweep angles of 0, 15, 30 and 45 degrees. They found out that increasing the sweep angle cavitation inception occurs at quite higher speeds and thus higher speeds of sub-cavitating flow could be attained. They showed that the sweep should alleviate the problem of erosion due to cavitation. Kuiper [2] improved the inception behavior of tip vortex cavitation on propellers. He did so by adding forward skew to the propeller and compared the result to those obtained from cases with no skew and backward skew. Ausoni et al. [3] investigated the hydrofoil roughness effects on von Kármán vortex shedding. They showed that with the help of a distributed roughness, the transition to turbulence is triggered at the leading edge, which reduces the span-wise non uniformities in the boundary layer transition process. Akbarzadeh et al. [4] studied the effect of blowing and suction on hydrodynamic behavior of sheet cavitating flows over hydrofoils. They found that by increasing the blowing amplitude and the width of jet, the lift and pressure drag coefficients are decreased. Coutier-Delgosha et al. [5] studied the effect of the surface roughness on the dynamics of sheet cavitation on a two-dimensional foil section. They presented that the roughness in the downstream end of the sheet cavity plays a major role in the arrangement of the cavitation cycle. Their results showed that the effect of roughness is a disorganization of the periodical shedding, characterized by much lower pressure fluctuations than previously. Danlos et. [6] investigated the effects of the surface condition of a venturi profile using visualization. They showed that the roughness surfaces can suppress the cloud cavitation shedding in a large range of cavitation numbers. The idea of using obstacles to generate cavitation artificially was used by Escaler et al. [7] In their work, they added large appendages to a hydrofoil to create intense cloud cavitation in order to accelerate erosion of the surface. Rhee and Kim [8] investigated numerical analysis of rudder cavitation in propeller slipstream and the development of a new rudder system aimed for lift augmentation and cavitation suppression. The new rudder system is equipped with cam devices which effectively close the gap between the horn/pintle and movable wing parts.

Most of the cavitation control investigations in previous works of researchers are attempted in super-cavitating regime. The behavior of super-cavitating and cavitating flow around a conical body of revolution with and without ventilation at several angles of attack was studied experimentally by Feng et al. [9]. Their experimental results showed that the drag of the model decreased with the presence of supercavity under the ventilation condition. Ji et al. [10] investigated both natural and ventilated cavitation numerically using the solver of a commercial CFD code CFX. They simulated the cavitating flow around an under-water vehicle under different cavitation conditions. They showed that the ventilated flow rate of the non-condensate gas influences the development of natural cavitation as well as ventilated cavitation. The vapor cavity is suppressed significantly by the gas cavity with the increase of the gas ventilation. Kadivar et al. [11] investigated super-cavitation flow over different 30, 45 and 60 degree wedge cavitators. They presented that the cavity length strongly depends on the velocity and wedge angle. They illustrated that the wedge angle of the cavitator affects significantly the shape and type of super-cavitation. Nesteruk [12] simulated the unsteady evolutions of the slender axisymmetric ventilated supercavity. He showed that the ventilation can increase and diminish the supercavity dimensions. Stability of steady and pulsating gas cavities was investigated in the case of the low gas injection rate. Javadpour et al. [13] investigated the numerical and experimental analysis of the ventilated super-cavitating flow around a cone cavitator. They showed that at constant rate of the ventilated air, with an increase of cavitation number from 0.15 to 0.25, the drag force drops by about 60 percent.

Careful review of the past works reveals that most of the previous investigations were focused on resolving cavitation and understanding its physics and despite of the importance of the cavitation control, it has been studied scare. The previous studies about the cavitation control methods which have attempted concentrated mostly on the control of the super-cavitation to reduce the drag reduction of the immersed body when it moves in water with high speed. According to the previous studies, it is well known that the control of cavitation on the hydrofoil has a significant influence on the destructive effects of cavitation such as drag reduction and vortex-induced vibration. We introduce in this work a method to control the cavitation which is of importance from many engineering points of view, in particular for marine engineering. The idea has adapted from vortex generators that are very common in boundary layer control around airfoils in aerospace engineering applications. This analogy is used to control boundary layer thickness and as a consequence the upper surface pressure distribution and to reduce the destructive effects of cavitation.

The main undesirable effects of cloud cavitation phenomenon is its unsteady and cyclic behavior. For this reason, the main purpose of this project is to introduce a method to control this unsteady and cyclic behavior. If we control the flow and create a condition, the bubble of cavitation is created artificially and it never disappears. This cavitating bubble can affect the whole processes of vaporization, bubble generation and bubble implosion, which occur at normal condition without any control. Achieving this goal is possible through inserting a type of micro vortex generator called cavitating-bubble generators (CGs) on the upper surface of the hydrofoil where it is expected that the cavitation is produced naturally, see Javadi et al [24]. The 2D schematic view of the hydrofoil using some inserted CGs on the suction side of hydrofoil and the 3D view of one CGs located on the hydrofoil are shown in Figure 1. Our investigations on the size of the CGs show that it should be small enough so that it does not have a significant effect on the hydrodynamics performance of the hydrofoil. Because of unpleasant side effects that may occur, the shape, the size and the location of the CGs are curtail and choosing L and H are based on running many test cases.

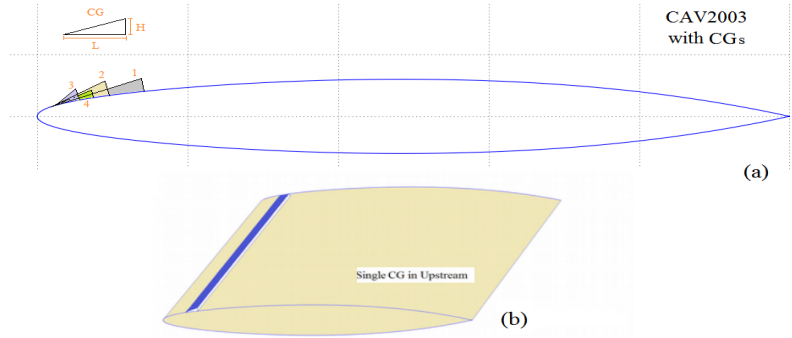

Figure 1: a) 2D view of some example CGs located on the hydrofoil with different sizes and angles, b) 3D view of one CGs located on the hydrofoil upstream

**Problem description**

Two-phase mixture model, based on the isotropic hypothesis for the fluid, along with the Partially-Averaged Navier-Stokes equations are employed. The two-phase mixture uses a local vapor volume fraction transport equation together with terms for the mass transfer rate between the two phases. The Schnerr and Sauer [14] cavitation model is chosen for numerical simulation of cavitation. The mass equation and momentum conservation can be written as following, respectively

$$\frac{\partial \rho}{\partial t} + \frac{\partial}{\partial x_i}(\rho u_i) = 0 \qquad (1)$$

$$\frac{\partial}{\partial t}(\rho u_i) + \frac{\partial}{\partial x_j}(\rho u_i u_j) = \rho f_i - \frac{\partial p}{\partial x_i} + \frac{\partial}{\partial x_j}\left[(\mu + \mu_t)\left(\frac{\partial u_i}{\partial x_j} + \frac{\partial u_j}{\partial x_i} - \frac{2}{3}\frac{\partial u_k}{\partial x_k}\delta_{ij}\right)\right] \qquad (2)$$

$$\rho = \alpha_v \rho_v + \alpha_l \rho_l \,, \qquad (3)$$

**PANS turbulence model**

In this paper PANS turbulence model is used as the turbulence model for the simulation of unsteady cavitating flow. The PANS model is a hybrid method RANS/DNS, which was first proposed by Girimaji [15]. The results of other publications using this turbulence model showed the improvement of the accuracy of numerical simulation, such as Ma et al. [16], Song and Park [17] and Lakshmipathy and Girimaji [18], . This turbulence approach is coupled with a mass transfer model which implements to the open source finite-volume package. The closure model was determined as a function of the ratio of the unresolved kinetic energy to the total kinetic energy k and the ratio of the dissipation to the total dissipation ε which are defined as the following

$$f_k = \frac{k_u}{k}, f_\varepsilon = \frac{\varepsilon_u}{\varepsilon} \tag{4}$$

$$\frac{\partial}{\partial t}(\rho k_u) + \frac{\partial}{\partial x_j}(\rho u_j k_u) = \frac{\partial}{\partial x_j}\left[\left(\mu + \frac{\mu_t}{\sigma_{k_u}}\right)\frac{\partial k_u}{\partial x_j}\right] + P_{k_u} - \rho \varepsilon_u, \tag{5}$$

$$\frac{\partial}{\partial t}(\rho \varepsilon_u) + \frac{\partial}{\partial x_j}(\rho u_j \varepsilon_u) = \frac{\partial}{\partial x_j}\left[\left(\mu + \frac{\mu_t}{\sigma_{\varepsilon_u}}\right)\frac{\partial \varepsilon_u}{\partial x_j}\right] + C_{\varepsilon 1} P_{k_u} \frac{\varepsilon_u}{k_u} - C^*_{\varepsilon 2} \rho \frac{\varepsilon_u^2}{k_u}, \tag{6}$$

$$C^*_{\varepsilon 2} = C_{\varepsilon 1} + \frac{f_k}{f_\varepsilon}(C_{\varepsilon 2} - C_{\varepsilon 1}) \tag{7}$$

$$\sigma_{k_u} = \sigma_k \frac{f_k^2}{f_\varepsilon}, \sigma_{\varepsilon_u} = \sigma_\varepsilon \frac{f_k^2}{f_\varepsilon} \tag{8}$$

**Numerical procedure**

In this study OpenFOAM [19] is used to solve the PANS equations in the cavitation flow. PANS is implemented into this finite-volume package. We use a modified interPhaseChangeFoam solver for simulation of cavitating flow. The time step is adjusted to ensure a maximum courant number about 1.0 in computational process. The pressure-velocity coupling is solved by a PIMPLE algorithm. The initial conditions and reference values for the simulation are chosen from the work by Delgosha et al. [20]. The effect of using three different mesh sizes on the time-averaged lift and drag coefficients and Strouhal number based on the chord are investigated. The simulations was performed using grid with 1,246,000 cells in the entire computational domain.. Figure 2 illustrates the computational domain and zoom view at the hydrofoil leading edge. The value of y + at the wall surface of hydrofoil is about 1.

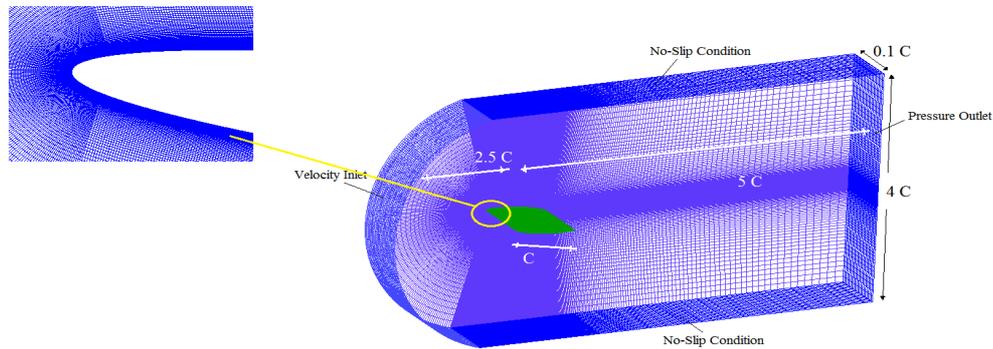

Figure 2: Computational domain and zoom view at the hydrofoil leading edge

**Results and Discussion**

Results of unsteady cloud cavitating flow over the CAV2003 benchmark hydrofoil with and without of the cavitation control are presented here. The geometry of this hydrofoil was offered in CAV2003 workshop [21].

Firstly we presented the unsteady cavitating flow around the hydrofoil without CGs to evaluate the numerical simulations based on experimental data. Acosta [22] was proposed different non-dimensional parameters such as maximum length of the attached part of the cavity to the chord $l/l_{ref}$, Strouhal number based on the chord $St_c$ and Strouhal numbers based on the cavity length $St_l$ according to the parameter $\sigma/2\alpha$ in his linearized theory of partial cavitation on flat plate hydrofoils. Two different types of partial cavity oscillations was meseared by Arndt et al. [23] in St. Anthony Falls Laboratory cavitation tunnel. They characterized these two cavitation dynamics by the parameter $St_c$, $St_l$ and $\sigma/2\alpha$ and found the limit of $\sigma/2\alpha = 4$ between these two types. They showed that for $\sigma/2\alpha > 4$, the flow instability of cloud cavitation flow mainly governed by the periodic reentrant-jet leading to $St_l = 0.3$. Furthermore, they reported that for $\sigma/2\alpha < 4$ other effects such as pressure wave propagation play also a significant role for the instability of cloud cavitation. Therefore 3D-dimensional calculations should be necessary in these cases with complex situations.

In our work the cavitation number at the outlet section was set to 0.8 and the dynamics of cloud cavitation is considered for the case $\sigma/2\alpha < 4$ as the same with Delgosha et al. [20] wherein the frequency of the oscillations for this case was reported about 6.5 Hz giving a period T = 0.15 s. The maximum length of the attached cavity to the chord $l/l_{ref}$ was presented 70 %. In Figure 3 one complete cycle of unsteady cavitation from our simulation was shown and was copmared with the experimental data. In this results the consecutive steps of the unsteady process, the cavity break-off and the convection of the vapor cloud was clearly observed. In our simulation the periodical unsteady behaviors characterized by values of Strouhal number $St_c = 0.11$ which is close to the numerical simulation and the experiments of Delgosha et al. with the Strouhal number of $St_c = 0.108$ and $St_c = 0.15$ respectively for the case without passive controller. The shedding frequency of cloud cavities for our case was observed f = 6.6 Hz.

The vapor volume fraction for the 3D cloud cavitation during one oscillation cycle using a proper manner of passive controller CGs was shown in Figure 4. These consecutive images show that the separation of the large cavity in the middle of the cavity region don't carried out using the passive controller. The cavity growth begins near the leading edge of the hydrofoil and increases gradually to a certain region of the hydrofoil surface which shows the maximum length of the attached cavity about the half of the hydrofoil. The maximum length of the attached cavity to the chord $l/l_{ref}$ is about 55 % for the cavitating flow using CGs which is much smaller than the maximum length of the attached cavity for the cloud cavitation without CGs. Fig 4 (e-h) shows a small changes of the cavity length at the closure region on the hydrofoil surface. Figure 4 also shows that the cavitating flow using proper CGs in cloud cavitation regime was changed to a sheet cavity with only oscillations in the downstream part and the large shedding of cloud cavitation has been suppressed.

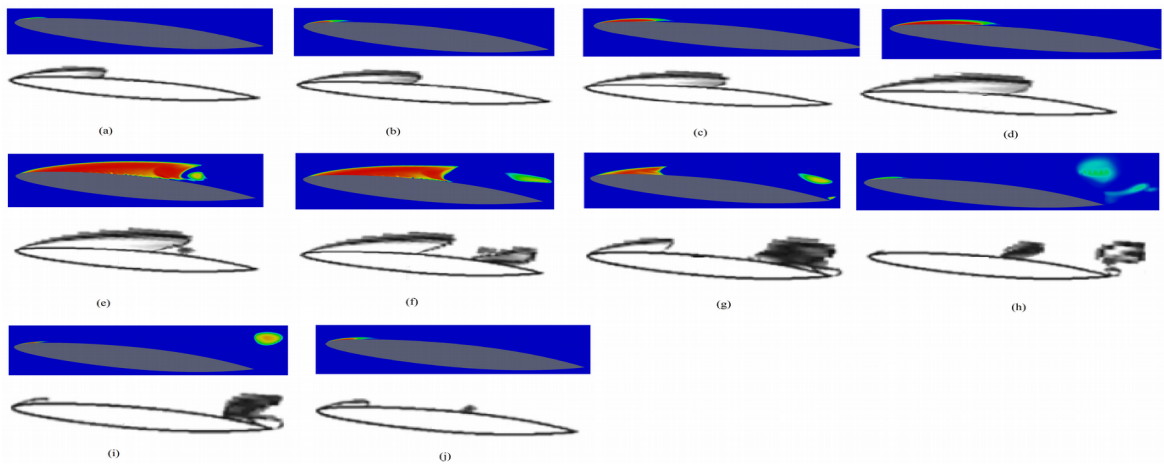

Figure 3: Instantaneous vapor volume fraction contours for the growth and collapse of the cavitation in one cycle, top) present work, bottom) Delgosha et al. (2007)

When the large number of vapor structures as bubble clusters and small-scale vortices reach the high pressure region on the surface of hydrofoil, they collapse and emit shock waves. This process can induce different destructive effects such as vibration, high wall-pressure peaks on the surface of hydrofoil and erosion. The influence of optimal CGs on the reduction of high-pressure peaks and stabilization of fluctuations in cloud cavitating flows was anlyzed in this work. The Figure 5 presents the wall-pressure fluctuations on suction surface of hydrofoil at different locations with and without using passive controller. As can be seen in this figure for hydrofoil without using CGs high wall-pressure peaks are captured during the cloud cavitation process.

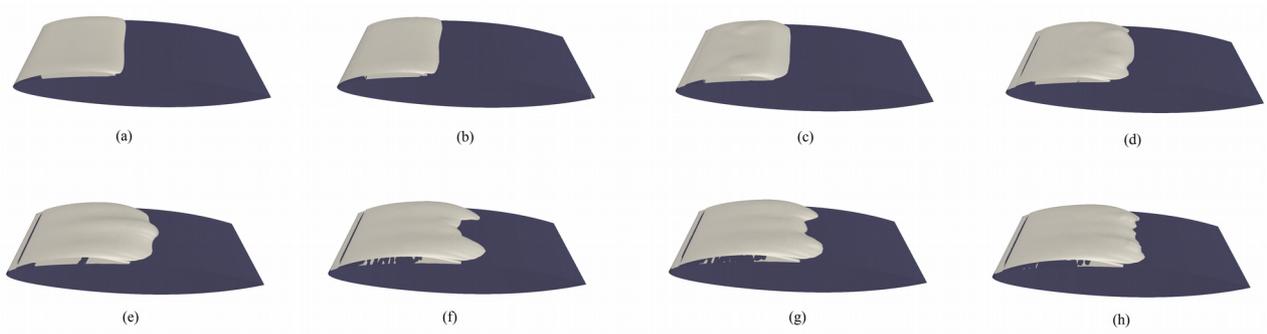

Figure 4: Vapor volume fraction iso-surface for the 3D cloud cavitation during one oscillation cycle with CGs

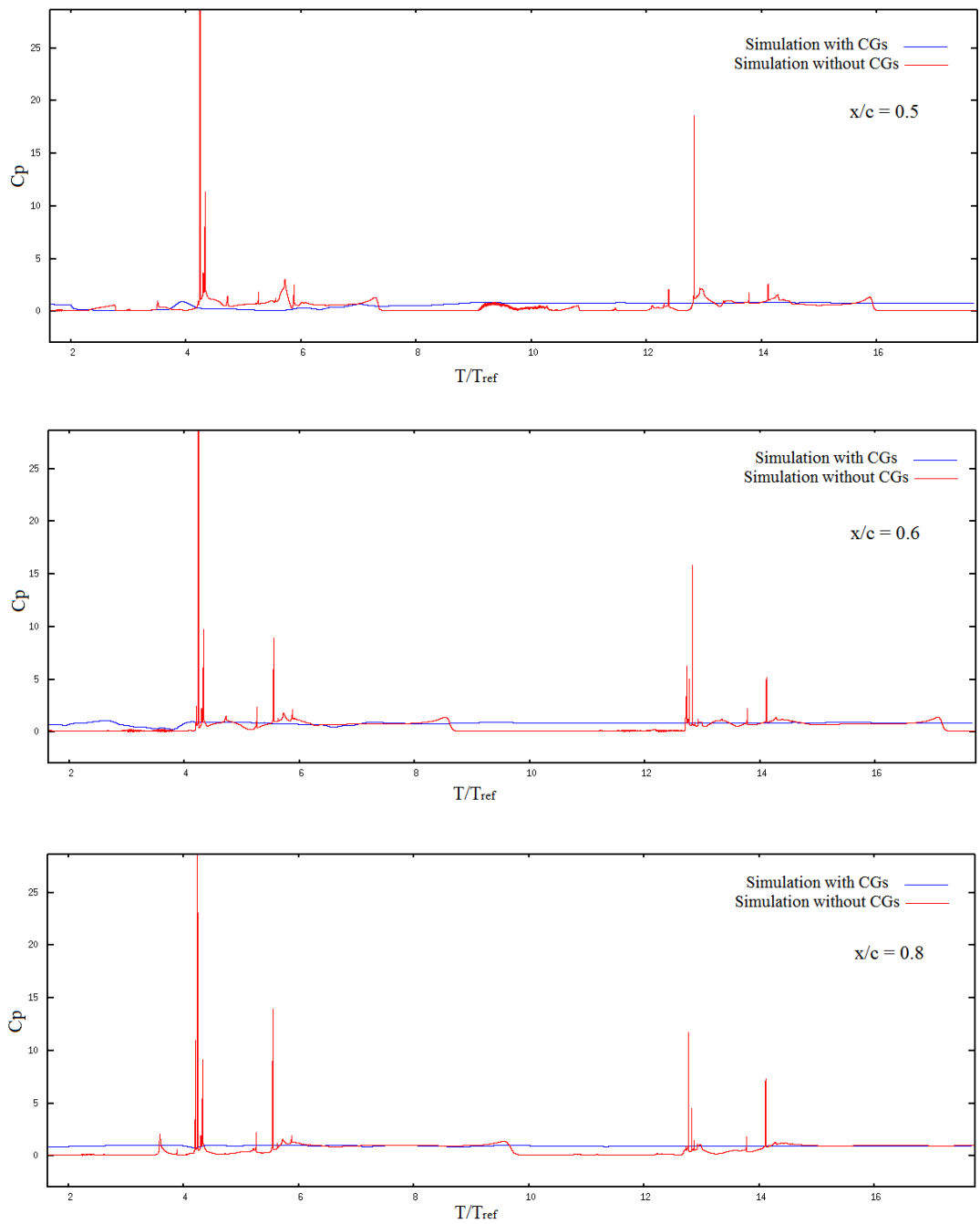

Figure 5: Wall pressure fluctuations on suction surface of hydrofoil at different locations x/c = 0.5, 0.6, 0.8

But for hydrofoil with using optimal CGs no pressure peaks are observed. The high peak values of the pressure fluctuations using optimal design of CGs are reduced significantly and the cavitating flow reaches a steady state situation which shows no high wall-pressure fluctuations and no periodical cavitation cycle. Therefore, this passive control method can be used to reduce the adverse effects of pressure which induced by high pressure peaks on the surface of hydrofoil and reduce the destructive effects of vibration on the immersed bodies.

Figure 6 shows fast Fourier transform (FFT) of lift progress on the hydrofoil with and without using passive controller. It can be seen from this figure that the several peaks are obtained with different magnitudes. For the case without using CGs four different peak frequencies $f_1$ = 6.6 Hz, $f_2$ = 12.4 Hz, $f_3$ = 19.13 Hz and $f_4$ = 25.8 Hz and for the case with using CGs two different peak frequencies $f_1$ = 10.23 Hz and $f_2$ = 16.8 Hz are shown. These values show that the dominant frequency for the case of using CGs is increased significantly in comparison with the simple one. But the amplitude of the dominant frequency using optimal CGs was reduced remarkably. These two aspects means that the cloud cavitating flow was reached a stable situation which has not the negative effects of cloud cavitation as before. The dominant frequency with the highest amplitude corresponding to the cavitation shedding events is considered to find the Strouhal number. For the case of using CGs the dominant frequency corresponds to Strouhal number values of $St_c$ = 0.17 and $St_l$ = 0.093 while for the case without CGs the Strouhal number values are about $St_c$ = 0.11 and $St_l$ = 0.077. As the figure shows that the second peak frequency f = 16.8 Hz of the simulation case with using CGs is also bigger than the peak frequency f = 12.4 Hz of the simulation without using passive controller.

For the cavitating flow using CGs the maximum length of the attached cavity was observed much smaller than the attached cavity for the simple hydrofoil in cloud cavitation regime. This means that the peak frequency corresponds to the shedding frequency of the cloud cavitation increases when the sheet cavity length decreases. The unstable regime of cloud cavitation which shows the large sheet cavity and large shedding of cloud cavitation could be suppressed using this passive control method as shown in Figure 6.

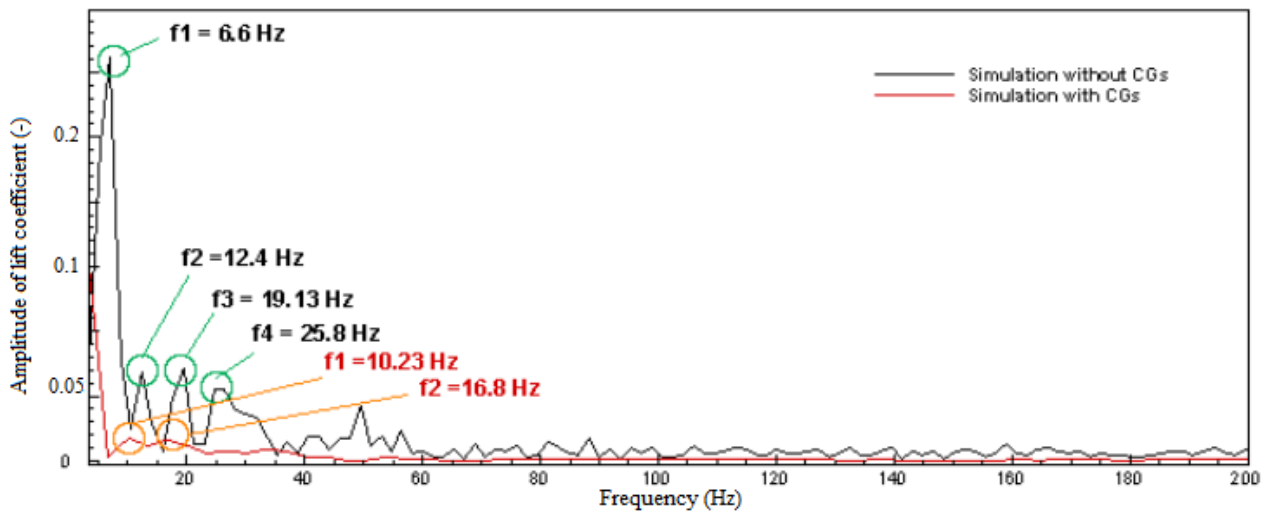

Figure 6: Frequency spectrum analysis of lift coefficient on hydrofoil CAV2003 with and without CGs

**Summary and conclusions**

In this work the effect of CGs as cavitation passive controller on the destructive effect of cloud cavitation such as unsteadiness and cycle behavior of the cloud cavitation and high wall-pressure peaks on the hydrofoil surface has been analyzed. Our results showed that the appropriate design of the CGs on the hydrofoil surface causes to the reduction in high-pressure peaks on the wall surface of the hydrofoil. The cyclic behavior of the unsteady cloud cavitation has been also suppressed using the proper CGs. In conclusion, using this method of passive control on the surface of immersed bodies such as hydrofoil, propeller and ruder which usually operate in cloud cavitation regime a significant reduction in vibration generation, unpleasant unsteady side force effects and erosion on the surface can be expected.